\documentclass{elsart}
\usepackage{epsfig}
\usepackage{amssymb}

\def\GeVc{{{\rm GeV}/c}}

\begin{document}
\begin{frontmatter}

\vbox{\normalsize%
\noindent%
\rightline{\hfill {\tt Belle Preprint 2002-9}}%
}
\vbox{\normalsize%
\noindent%
\rightline{\hfill {\tt KEK Preprint 2002-12}}%
}
\vspace{0.7cm}

\title{
A Detailed Monte-Carlo Simulation for the Belle TOF System}

\author{ J.~W.~Nam, Y.~I.~Choi, D.~W.~Kim and J.~H.~Kim } \address{Sungkyunkwan 
University, Suwon}

\author{ B.~C.~K.~Casey, M.~Jones, S.~L.~Olsen, M.~Peters,}
\author{ J.~L.~Rodriguez, G. Varner  and Y.~Zheng} \address{University of Hawaii, Honolulu HI}

\author{ N.~Gabyshev, H.~Kichimi and J.~Yashima } \address{ High Energy Accelera
tor Research Organization (KEK), Tsukuba}

\author{ J.~Zhang } \address{University of Tsukuba, Tsukuba}

\author{ T.~H.~Kim and Y.~J.~Kwon } \address{Yonsei University, Seoul}

\vspace{0.7cm}

\normalsize
\begin{abstract}
We have developed a detailed Monte Carlo simulation program for
the Belle TOF system. Based on GEANT simulation, it takes account of
all physics processes in the TOF scintillation counters and readout
electronics.
The simulation reproduces very well the performance of the Belle TOF system,
including the $dE/dx$ response, the time walk effect, the time
resolution, and the hit
efficiency due to beam background. In this report, we will describe the
 Belle TOF simulation program in detail.

\end{abstract}

\end{frontmatter}

\clearpage

\section{Introduction}
A time of flight (TOF) detector\cite{TOFNIM,NIM} system using plastic
scintillation counters is very powerful for particle identification
(PID) in $e^+ e^-$ collider detectors~\cite{NIM,kekb}. A TOF system
with $100~\rm{ps}$ time resolution is effective for particle momenta below about
$1.2~\GeVc$, which encompasses 90\% of the particles produced in
$\Upsilon (4S)$ decays. It can provide clean and efficient flavor tagging
for CP violation studies in B meson decays in a B-factory experiment.

A reliable Monte Carlo (MC) simulator is a good tool to understand the
detector system and is important for minimizing experimental systematic
uncertainties for physics analysis. Although one simple
simulation method is a parametrization based on real data, it
doesn't provide precise simulation for complex experimental
situations, such as multi-track hits, hadronic interactions
in the TOF counter, and shower leakage from the calorimeter outside
of the TOF system.

We have developed a full detector simulator for the TOF system. Based on
GEANT3~\cite{GEANT}, it includes the simulation of all physics
processes involving scintillation light production and its propagation in
the TOF counters, the production of signal pulses in the photomultiplier tubes
(PMT), the electronics process in the discriminator and the charge to
time converter (QTC), and the effects of beam background.

In this report, we describe the Belle TOF detector
simulation. In section ~\ref{sec:tof}, we briefly introduce the Belle TOF
system. In section ~\ref{sec:simulation} we describe the simulation model
and the main algorithm for each process. Finally, we compare the
simulation results with experimental data in section ~\ref{sec:result}.
Section ~\ref{sec:conclusion} contains conclusions.

\section{TOF system in the Belle detector}
\label{sec:tof}

\begin{figure}[ht]
\centerline{
\epsfysize=4.0in
\epsfbox{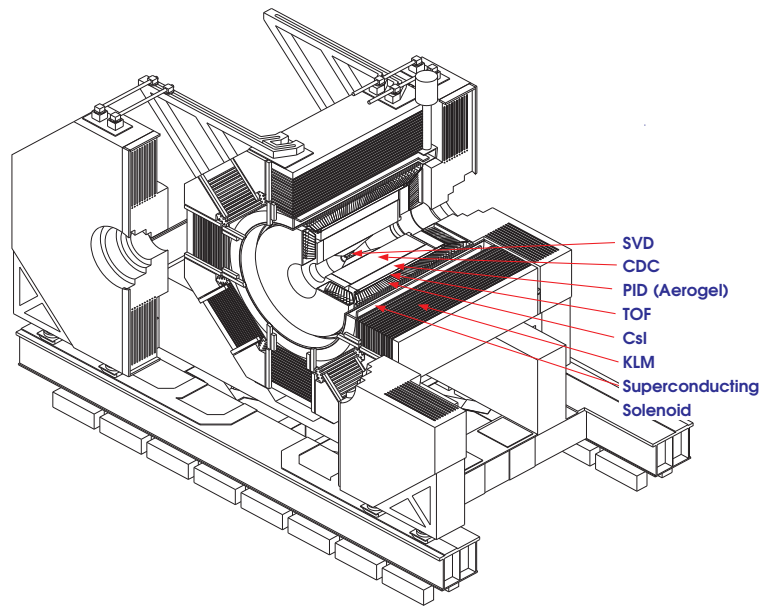}}
\caption{A cut-away view of the Belle detector.}
\label{fig:belle}
\end{figure}

The Belle detector is a large solid-angle spectrometer based on a $1.5~\rm{T}$
superconducting solenoid magnet; a detailed description can be found
elsewhere~\cite{NIM}. Figure~\ref{fig:belle} shows the Belle detector.

\begin{figure}[ht]
\centerline{
\epsfysize=2.7in
\epsfbox{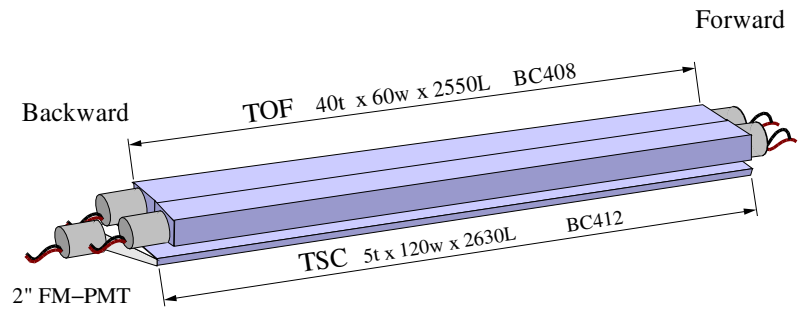}}
\caption{A TOF/TSC module.}
\label{fig:tof_counter}
\end{figure}

The TOF system consists of 64 TOF/TSC modules located at a radius of $1.2~\rm{m}$ from the interaction point, which cover a polar angle of $33.7^\circ \sim
120.8^\circ$. The design of one TOF module is shown in
Figure~\ref{fig:tof_counter}. A module consists of two TOF counters of $4~\rm{cm}$
thickness, viewed at both ends by PMTs, and one $0.5~\rm{cm}$ thick TSC counter,
viewed only at the backward (i.e. large angle) end.

In order to get sufficient gain in a $1.5~\rm{T}$ field, a fine mesh
photomultiplier (FM-PMT)~\cite{FMPMT} is used. This tube has 24
fine mesh dynodes of $2000~\rm{mesh/inch}$, and its size is $8~\rm{cm}$ in length and 5.08
cm in diameter. Its bialkali photocathode covers 50\% of the end area
of each TOF counter. The PMT gain is adjusted to be about $1 \times
10^6$,
which requires a high voltage of about $2100~\rm{V}$ in the $1.5~\rm{T}$ field.
The FM-PMTs are attached
to the TOF counter ends with an air gap of $\sim~0.1~\rm{mm}$ that
selectively passes early arrival photons and reduces the PMT anode
current. For the TSC counters, 
the tubes are glued to light guides that are attached to the sintillator.
Table~\ref{tb:TOFPMT} lists the characteristic
parameters of the TOF scintillator and the FM-PMT used in this simulation.

\begin{figure}[ht]
\centerline{
\epsfysize=2.5in
\epsfbox{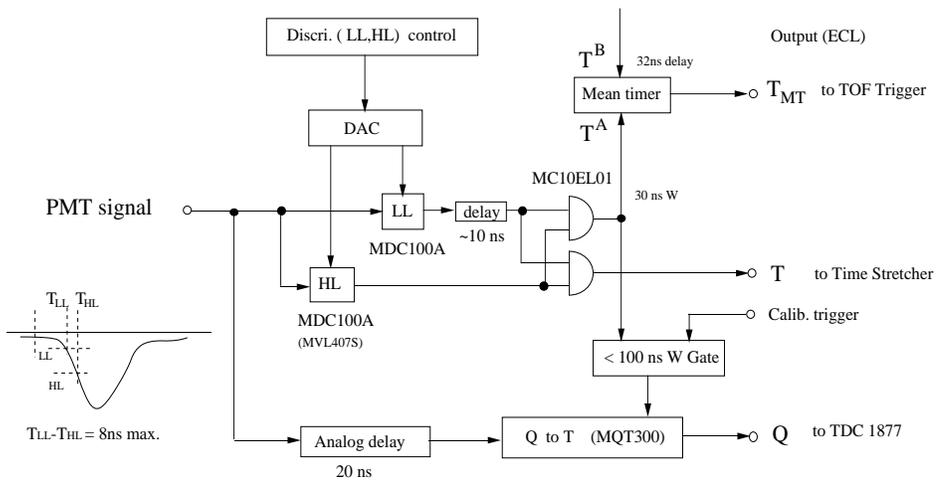}}
\caption{A block diagram of the TOF front end electronics (TOFFEE).}
\label{fig:toffee}
\end{figure}

Figure~\ref{fig:toffee} shows a block diagram of the TOF front end
electronics (TOFFEE).
Each PMT signal is split in two. One half is sent to a
charge-to-time-converter (QTC) for charge measurement. The other half  goes to
generate two output signals by application of different threshold
levels. The high level (HL) signal is used to
provide a gate to the QTC, and the low level (LL) signal is fed to a time stretcher
(TS) for time measurement.

\section{Simulation procedure}

\label{sec:simulation}

\begin{figure}[ht]
\centerline{
\epsfysize=5.0in
\epsfbox{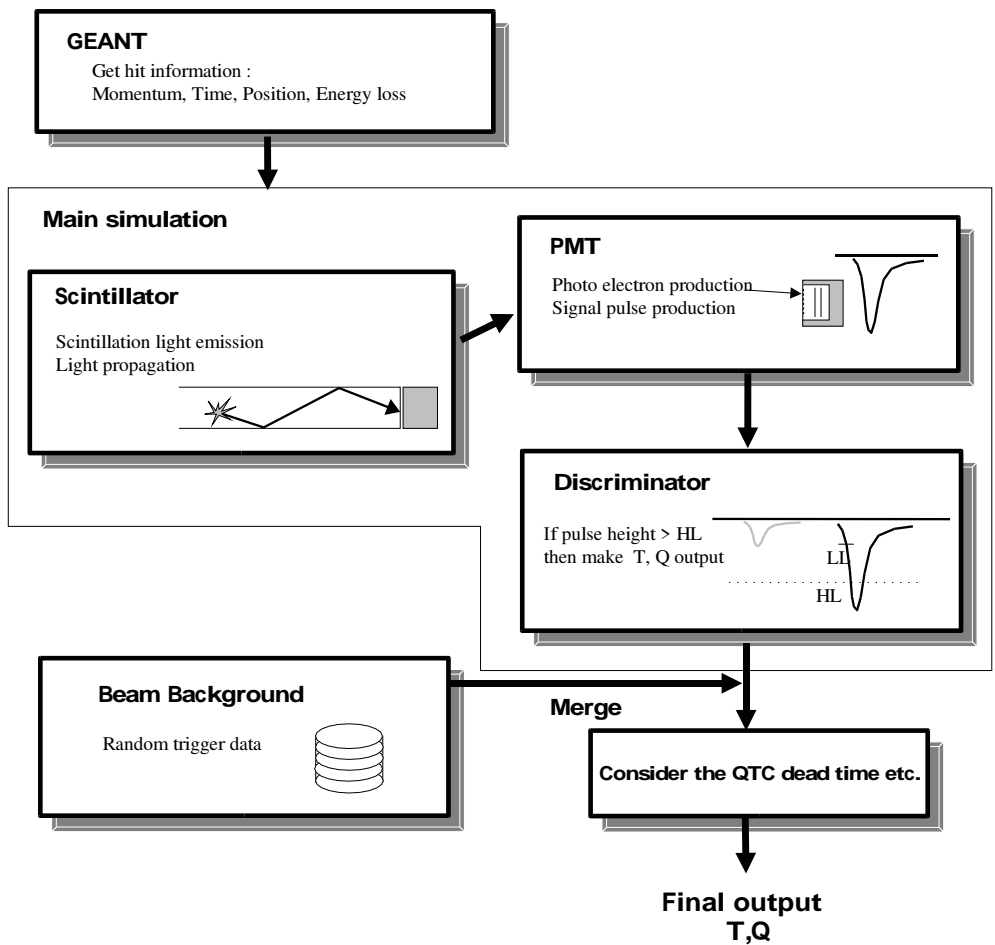}}
\caption{A block diagram of the simulation program.}
\label{fig:outline}
\end{figure}

Figure~\ref{fig:outline} shows the outline of the Belle TOF simulation
program. It consists of the GEANT part, the main TOF simulation part
including the scintillator, the PMT and discriminator processes, and the
beam background merge.

GEANT provides the hit information for a particle passing through the
scintillator, including the four vector, hit position, time, and
energy loss.

Using the hit information from GEANT, scintillation light is
produced. The light is then propagated in the TOF scintillator until it arrives at
a PMT and makes a signal pulse. Finally, a pulse larger than the HL threshold
makes a gate to measure the charge and timing of the signal. The main
detector characteristics, such as the time resolution, are simulated in
this part.

In order to take into account the effect of beam background, real
data taken with random triggers, which represent the beam background, are
merged into the MC output.  The inefficiency due to the dead time of
the QTC is simulated here.

\subsection{Geant simulation up to the TOF counters}

The geometrical and material configuration of the Belle detector is
implemented in the Belle GEANT full simulator~\cite{NIM}. Not only all
sub-detector components, but also materials for the support, cooling
structures, and readout cables are included. These allow realistic
simulation of a primary particle that goes through the inner
sub-detector
(and materials) up to the TOF scintillator and of secondary
particles that are produced in the inner and outer materials.

The material of the TOF/TSC scintillator is defined as a mixture of
hydrogen and carbon. The atomic ratio is assumed to be 1:1. The density
of this material is 1.032.

\begin{figure}[ht]
\centerline{
\epsfysize=3.5in
\epsfbox{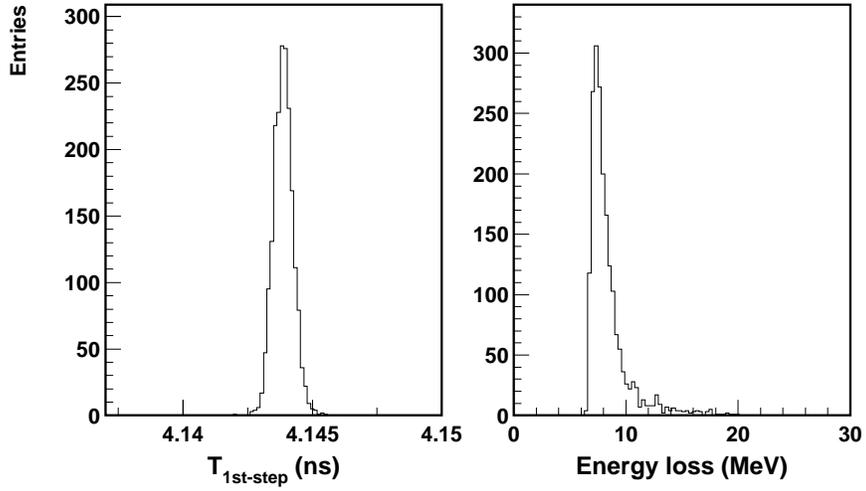}}
\caption{The first-step time (left panel) and energy loss (right panel) distributions.}
\label{fig:step}
\end{figure}

The GEANT hit information for a particle traversing the scintillator
volume is
recorded at discrete steps along the particle trajectory. For a
particle with a given energy, the step size is automatically calculated by
GEANT depending on particle species and on the characteristics of the
current medium~\cite{GEANT}. Figure~\ref{fig:step} shows the time at the
first step and energy loss distributions for $2.0~\GeVc$ $\mu^-$ which
enter perpendicular to the TOF counter.

\subsection{Scintillation process}
Using the step information from GEANT, we reconstruct a
continuous path connecting each step in the scintillator volume.
Scintillation light is uniformly emitted along this path. The time
$t_{traj}$ on this path for each light emission event is calculated by using
the four vector of the particle.

\begin{figure}[th]
\centerline{
\epsfysize=3.5in
\epsfbox{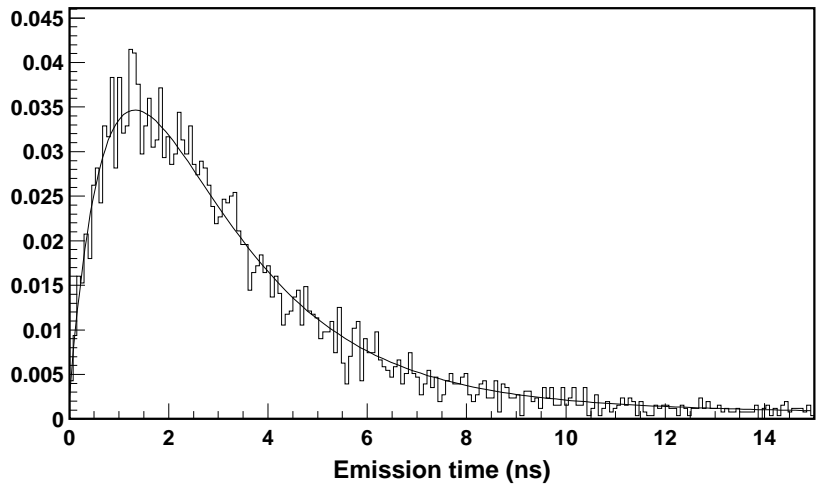}}
\caption{The emission time probability for scintillation light. The line is the function, the histogram is the simulated result.}
\label{fig:scintillation}
\end{figure}

The number of photons is proportional to the energy loss of the particle,
at a rate
assumed to be one photon per $100~{\rm eV}$ \cite{Clark}. The angular
distribution of the emitted light is isotropic.
The time profile of scintillation emission is simulated by using the
emission
time ($t_{emit}$) probability function~\cite{scintillation},

\begin{equation}
E(t_{emit}) = \frac{1}{1+R} \left( \frac{e^{-t_{emit}/\tau_2} -
e^{-t_{emit}/\tau_1}}{\tau_2-\tau_1} + \frac{R}{\tau_3}
e^{-t_{emit}/\tau_3}\right)
\end{equation}
where $\tau_1$ and $\tau_2$ are two fast decay time constants,
$\tau_3$ is the decay time of the slow component, and R is the ratio of the
slow to fast components. Since the signal timing is determined at the leading
edge in the discriminator, $\tau_1$ and $\tau_2$ are the most important
parameters for TOF simulation. We use
$\tau_1=0.9~{\rm ns}$ and $\tau_2=2.1~{\rm ns}$; these values are chosen
so that
this function has the same rise time as the BC408 scintillator light
output. The values $\tau_3=14.2~{\rm ns}$ and $R=0.27$ are given in
reference~\cite{scintillation}. Figure~\ref{fig:scintillation} shows the
$E(t_{emit})$ distribution.

\subsection{Light Propagation}

\begin{figure}[ht]
\centerline{
\epsfysize=5.5in
\epsfbox{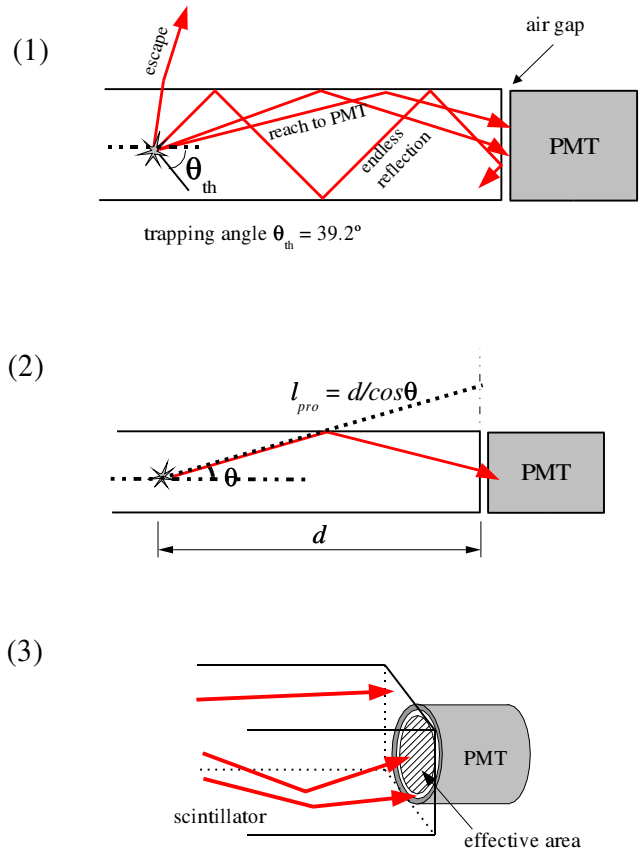}}
\caption{The light propagation in a scintillator counter.}
\label{fig:propagation}
\end{figure}

The scintillation light is propagated by total internal reflection
and arrives at the PMTs located at the ends of the scintillator.
We assume that light which has an emission angle within the trapping
angle ($\theta_{trap}$) can traverse the air gap to reach the PMT, while
the remaining light
escapes from the scintillator volume without total reflection or is
lost during the endless total reflection between the two ends of the TOF
scintillator (see Figure~\ref{fig:propagation}-(1)). The angle $\theta_{trap}$
is given by
\begin{equation}
\theta_{trap} = sin^{-1} \left( {\frac{n_{air}}{n_{scint}}} \right),
\end{equation}
where $n_{air}$ and $n_{scint}$ are the refractive indices of the
air and scintillator, respectively.

\begin{figure}[ht]
\centerline{
\epsfysize=3.5in
\epsfbox{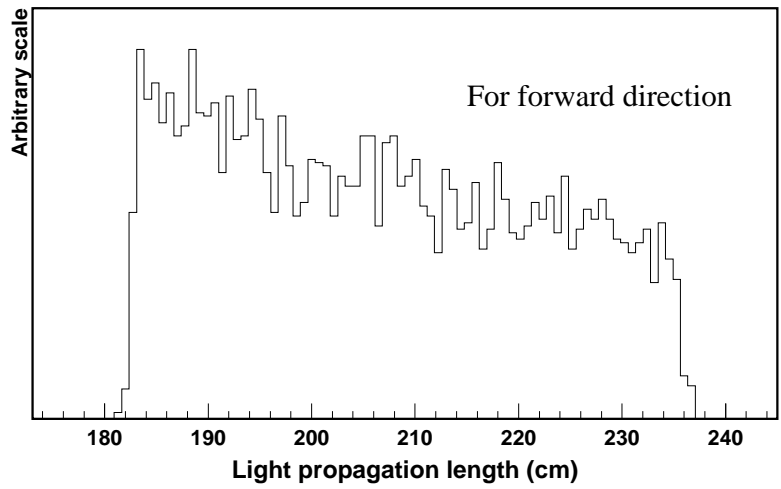}}
\caption{The light propagation length distribution.}
\label{fig:lightlength}
\end{figure}

By assuming  perfect internal reflection on each of the four surfaces of the
TOF
counter, the light propagation length in the counter is calculated (see
Figure~\ref{fig:propagation}-(2)) as
\begin{equation}
l_{pro} = \frac{d}{cos \theta},
\end{equation}
where $d$ is the distance from the light emission position to the end surface
of the scintillator and $\theta$ is the polar angle of the emitted
light. This technique effectively reduces the CPU time, while the result
is equivalent to a full simulation of light propagation.
Figure~\ref{fig:lightlength} shows the path length distribution of
light which is produced at $z=0$ and reaches the forward PMT.

The light propagation time in the scintillator is obtained by
\begin{equation}
t_{pro} = \frac {l_{pro} }{ v_{scint}},
\end{equation}
where $v_{scint}=c/n_{scint}$ is the velocity of light in the
scintillator.

Attenuation in the TOF counter is simulated according to
\begin{equation}
R(l_{pro}) = e^{-l_{pro}/\lambda_{pro}}
\end{equation}
where $\lambda_{pro}$ is the attenuation length in the scintillator
material. We
use $\lambda_{pro}=400~{\rm cm}$, which is measured with real data.

The mismatch of the rectangular cross section of the TOF counter to the
round PMT results in light loss (see
Figure~\ref{fig:propagation}-(3)). In the simulation, the number of
photons is reduced by the ratio of the TOF end area to the effective
area of the PMT photocathode.

\subsection{PMT response}

We consider the angular response ($R_{angle}$) of PMT sensitivity, which is a dependence on the angle of incident light on the photocathode~\cite{angular},

\begin{equation}
R_{angle} = cos \theta _{i},
\end{equation}

where $\theta_{i}$ is the light incident angle at the PMT front glass.
We find 0.66 for the efficiency averaged over all arrival directions at
the PMT. The number of photons is reduced by this factor.

At the PMT photocathode, photoelectrons are produced with a
probability which is called the quantum efficiency. The average quantum
efficiency of the photocathode is calculated to be 0.21 for the spectrum
from the BC408 scintillator.
Also, the number of photoelectrons is reduced by a factor of 0.6, called
the collection factor, due to the fine mesh dynode structure. We reduce the number of generated photoelectrons by these factors.

\begin{figure}[ht]
\centerline{
\epsfysize=5.0in
\epsfbox{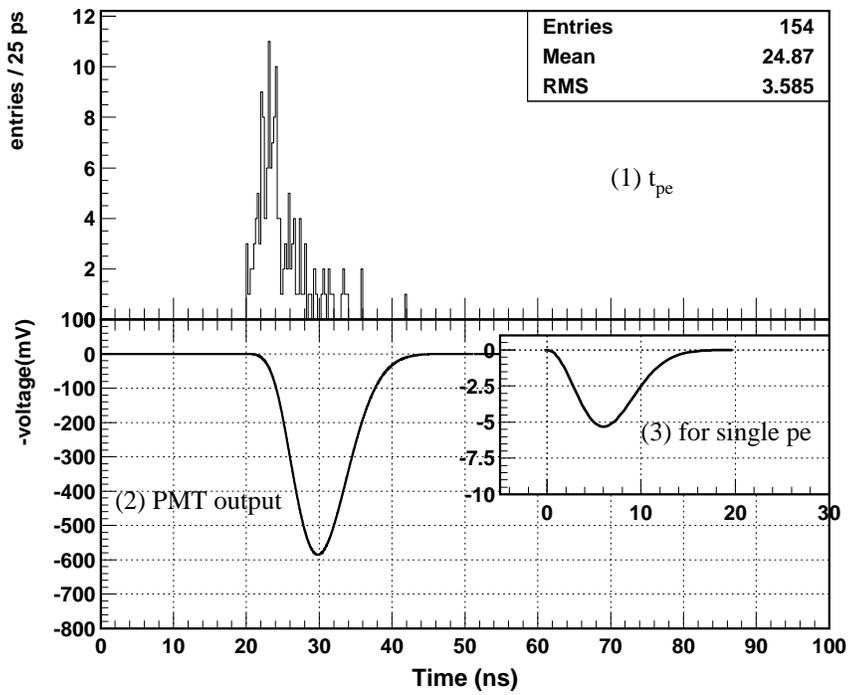}}
\caption{(1) The time distribution of the photoelectrons, (2) the PMT pulse output (with negative sign), and (3) the PMT response
for a single photoelectron.}
\label{fig:time_and_pulse}
\end{figure}

The transit time of a single photoelectron signal in a FM-PMT, $t_{TT}$,
is modeled by a transit time(TT) of $5.5~\rm{ns}$ and a r.m.s. transit
time spread (TTS) of $0.31~\rm{ns}$, which are measured for the R6680
FM-PMT.
By adding the smeared transit time to the photon timing information
described in the previous section, the pivot time of single
photoelectron signal at the anode is obtained,

\begin{equation}
t_{pe} = t_{traj} + t_{emit} +  t_{pro} + t_{TT}.
\end{equation}
Figure~\ref{fig:time_and_pulse}-(1) shows
the time distribution of the photoelectrons at the anode for a
typical event, a $2.0~\GeVc$ $\mu^-$ incident on a TOF counter at
$z=0$.

To simulate the time walk effect due to the leading edge discrimination
scheme used for the Belle TOF, a time response function $v_i(t)$ for
a single photoelectron is introduced. Using this response function,
the PMT output signal is obtained by summing all the
single photoelectron signals in the PMT for an event,
\begin{equation}
V_{PMT}(t) = \sum_{i=1} ^{n_{pe}} v_i(t),
\end{equation}
where
\begin{equation}
v_i(t) = G  C_e  \frac{t^2 e^{ -t^2/\tau^2}}{\int t^2 e^{ -t^2/\tau^2} dt
}.
\end{equation}
Here $G$ is the gain of PMT, $C_e$ is the conversion factor for charge to
voltage, and $\tau$ is the time constant. We use
$\tau=\rm 6~ns$ so that the rise time of this function is the same
as that of an R6680 PMT, and $25~\rm{ps}$ time bin size which is the same
as the least count of the Belle TOF readout system~\cite{TOFNIM}.

Figure~\ref{fig:time_and_pulse}-(2) shows a typical PMT output signal
for the same event as shown in Figure~\ref{fig:time_and_pulse}-(1).
The simulated signal is consistent with actual signals observed in the
Belle TOF counters in both pulse height and rise time.

\subsection{Simulation of readout electronics}
Each PMT signal is examined by readout electronics using double
thresholds.
A signal larger than the high level (HL) threshold provides a gate to
measure the signal timing (TDC) which is given at the moment when the
PMT pulse crosses the low level
(LL) threshold in the discriminator simulation. We use the same LL threshold
as the real experiment ($50~\rm{mV}$). A HL value 50\% higher than that in
the real experiment ($150~\rm{mV}$) is used for fine tuning of the hit
inefficiency, which is caused by PMT-to-PMT gain variation in the real
experiment.

\begin{figure}[th]
\centerline{
\epsfysize=5.5in
\epsfbox{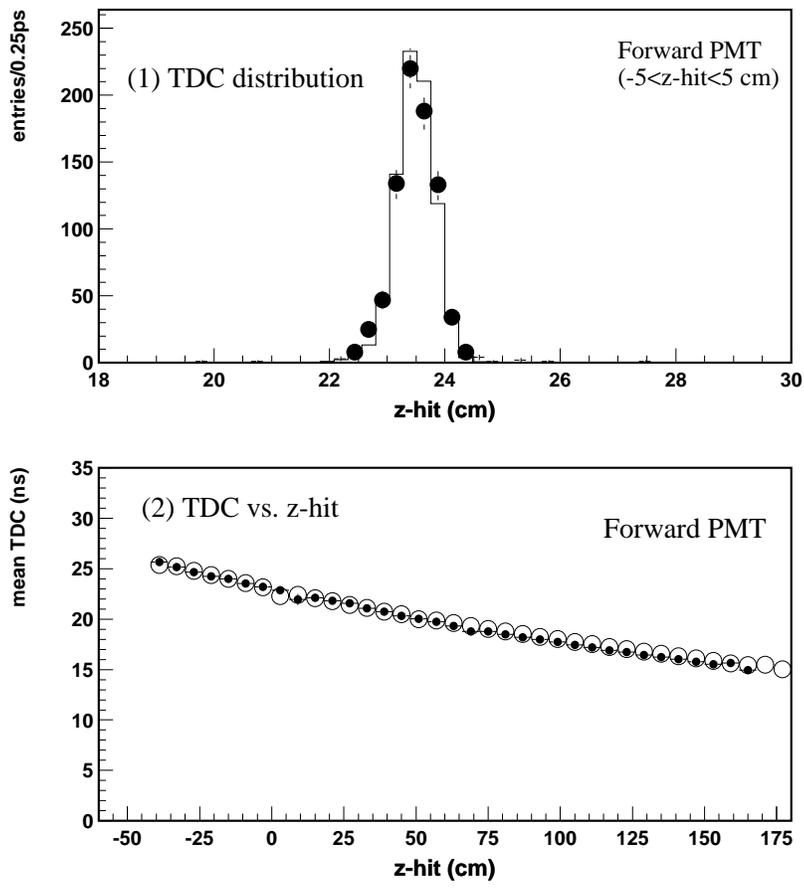}}
\caption{(1) The TDC distribution for -5$<z$-hit$<$5 cm, (2) the TDC
distribution as a function of $z$-hit. The solid circles are data; the
histogram and open circles are MC.}
\label{fig:tdc}
\end{figure}

Finally, we introduce a timing uncertainty, parametrized by a Gaussian
with a width of $45~\rm{ps}$, 
which is due to
the uncertainty of the beam crossing time ($t=0$) determination, the time
jitter caused by electronic noise in the discriminators, and the time
jitter in the TS readout.
Figure~\ref{fig:tdc}-(1) shows the TDC distribution for perpendicular
incidence on a TOF counter, and Figure~\ref{fig:tdc}-(2) shows TDC values
as a function of particle hit position ($z$-hit) along the counter.
The $z$-hit value is obtained by extrapolating the particle's trajectory
using tracking information from the central drift
chamber (CDC) ~\cite{NIM}. The filled circles in Figure~\ref{fig:tdc}
are data and the open
circles are MC. There is good agreement.

\begin{figure}[th]
\centerline{
\epsfysize=5.5in
\epsfbox{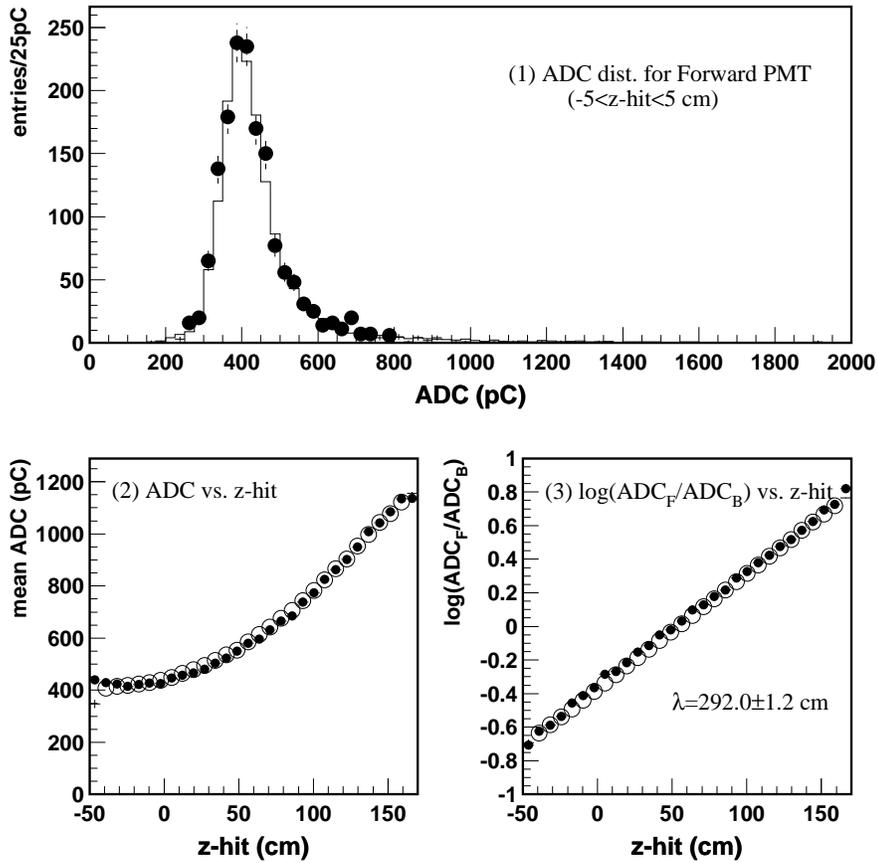}}
\caption{(1) The ADC output distribution. (2) The mean ADC value as a
function of $z$-hit. (3) The dependence of $log(ADC_F/ADC_B)$ on $z$-hit.  The solid circles are data; the histogram and open circles
are MC.}
\label{fig:adc}
\end{figure}

The charge output (ADC) is given by the time integral of the
$V_{PMT}(t)$ divided by the impedance of the PMT circuit. The width of the
simulated charge distribution is narrower than that of the real data.
Possible sources of the difference include
statistical gain variations of the PMTs~\cite{ACC} (the
simulation
assumes a constant gain for a single photoelectron), 
gain variation among individual PMTs, 
and the imperfection of the simplified model of
light propagation. To better represent the data, we adjust the
charge
distribution by introducing a scaled Poisson distribution,

\begin{equation}
{\rm ADC} =   Q_{orig} \left( S_{w} \frac{\left( N \left(n_{pe} \right)-n_{pe} \right)}{n_{pe}} + 1 \right),
\end{equation}
where ADC is the corrected charge, $S_{w}$ is the width scale parameter,
$N(n_{pe})$ is a random number given by a Poisson distribution generator
which has mean $n_{pe}$ and $Q_{orig}$ is the original charge from the
$V_{PMT}(t)$ integration.  In a comparison between $Q_{orig}$ and real
data, we found
that a value $S_{w}=2$ gives the best agreement.
Figure~\ref{fig:adc}-(1) shows the resulting charge distribution compared
with the real data. Figure~\ref{fig:adc}-(2) shows the mean ADC
value as a function of the $z$-hit position. All points are in
good agreement. By fitting $\log(ADC_F/ADC_B)$ as a function
of $z$-hit (see Figure~\ref{fig:adc}-(3)) to a linear function, we
obtain good agreement for the effective light attenuation length
($\lambda$) along the $z$ direction~\cite{aging1}.

\begin{figure}[th]
\centerline{
\epsfysize=4.0in
\epsfbox{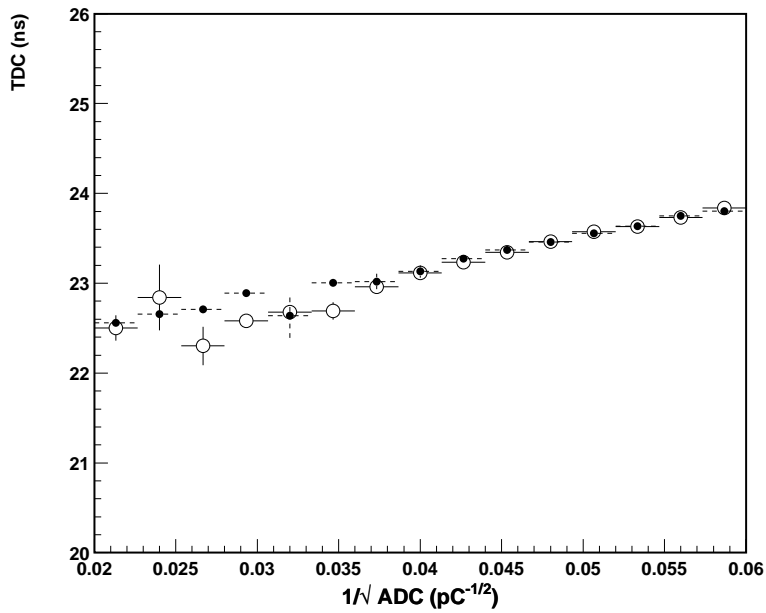}}
\caption{The variation of the TDC output {\it vs} $1/\sqrt{ADC}$ for -5$<z$-hit$<$5 cm. The solid circles are
data and the open circles are MC. The TDC and ADC values are for the forward PMTs.}
\label{fig:twc}
\end{figure}

One of the main parameters in the TOF calibration and reconstruction is a
time walk correction coefficient that is used to remove a pulse height
dependence~\cite{TOFNIM} of the digitized time. It is the coefficient of
a  $1/\sqrt{ADC}$ term, and is sensitive to the rise time of the PMT signal.
Figure~\ref{fig:twc} shows the (uncorrected) TDC versus $1/\sqrt{ADC}$
for data and MC. They are in good agreement.

\subsection{Beam Background Simulation}
The effect of the beam background is simulated  by  merging beam
background data
into the MC signal output.  The background data are taken  by  random
triggers during the real experiment. The typical hit rate of a single
TOF PMT is $15 \sim 25~\rm{kHz}$. Because of its greater
scintillator width, the TSC counter hit rate is more than 2 times higher
than the TOF rate.  The hit rates are expected to increase in the
higher luminosity environment in the near future.

\begin{figure}[th]
\centerline{
\epsfysize=5.5in
\epsfbox{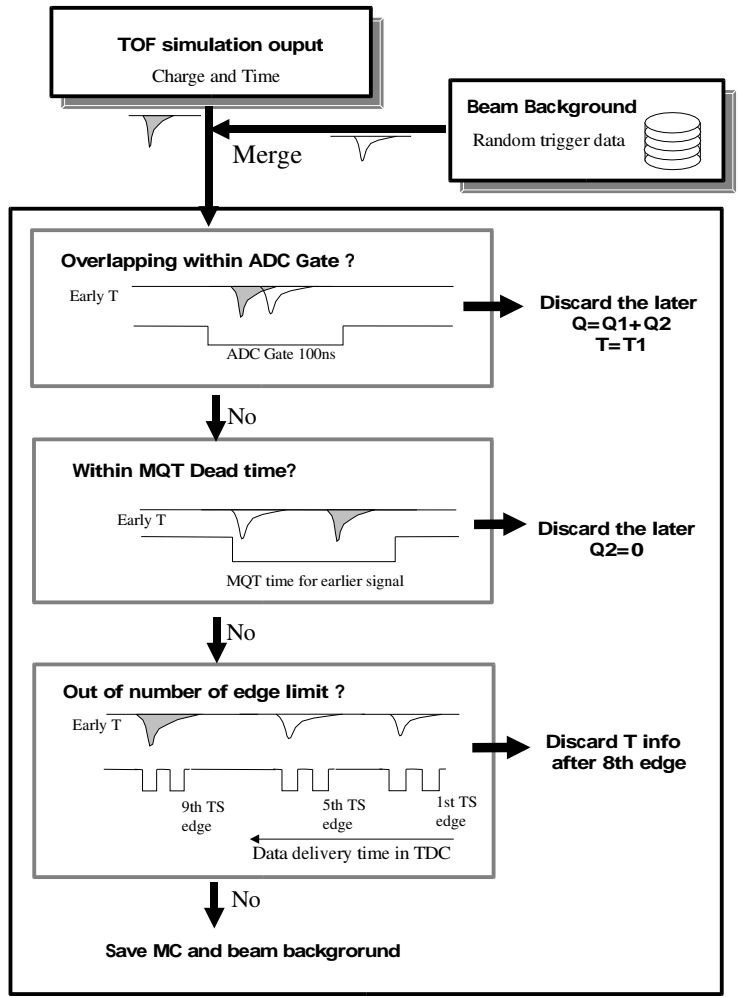}}
\caption{A flow chart for the merger of beam background with simulated events.}
\label{fig:addbg}
\end{figure}

Figure~\ref{fig:addbg} presents  a flow chart  of the  beam background
merging.  For each PMT, MC and background are merged on an
event-by-event basis. First, we consider the overlapping of two PMT
signals. According to the algorithm in the readout electronics, if two
signals
are closer than the ADC gate  width ($100~\rm{ns}$), we take only the earlier
signal TDC
and the charge  of the later signal  is  added to the earlier
one. Otherwise,
both signals are recorded.  Then the QTC dead time and  the number
of edges are taken into account.

The major effect of the beam  background is to increase the inefficiency
caused by the QTC dead time, due to signal conversion and recovery times.
It is simulated by discarding the later
TDC signal in the case that the MC and beam background times are
closer than the QTC dead time, which is found to be $0.55~\rm{ns}$ per ADC count (Each ADC count corresponds to a charge of $0.25~\rm{pC}$.) in real data.

Another source of inefficiency is the limit on the number of records
in the TDC1877 system~\cite{NIM}. The limit is 8 edge records for each
PMT in an event.
Because the TS system produces 4 edges for one PMT signal~\cite{TOFNIM},
a maximum of 2 PMT signals can be recorded for an event. In our
simulation, we count the number of edges for the individual TDC channels,
and remove the signal(s) after the 8th edge in the merged data. Since we
don't use the TS for the TSC counters, this inefficiency is not present
for them.

\begin{figure}[th]
\centerline{
\epsfysize=4.5in
\epsfbox{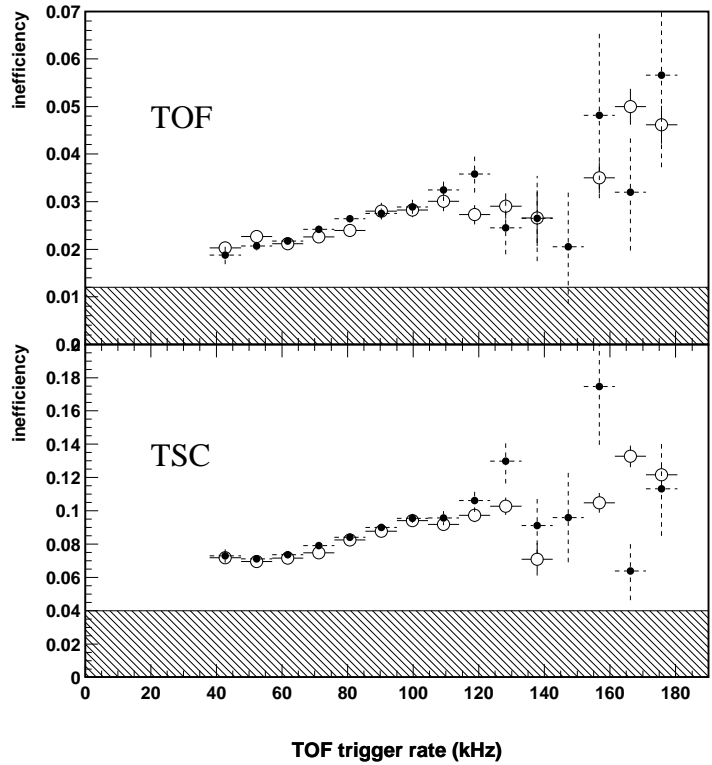}}
\caption{The hit inefficiency as a function of the TOF rate. The solid circles
are data and the open circles are MC.}
\label{fig:tofineff}
\end{figure}

Figure~\ref{fig:tofineff} shows the TOF and TSC hit inefficiencies
in  $\mu$-pair events $(e^+e^-\rightarrow \mu^+\mu^-)$ as a function of the TOF trigger rate, which
is defined as the rate for at least one TOF-TSC coincidence among
the 64 TSC counters.  The TOF PMT singles rate is about $25\%$ of
this TOF trigger rate, and the TSC singles rate is about two times
the TOF singles rate.  Here the hit
inefficiency is defined as the fraction of muons predicted to hit a TOF
counter for which there is no associated counter signal.
In the figure, the solid circles are for data
and the open circles are for MC with beam background. The hatched part
indicates the inefficiency given by the original MC without beam
background. This
intrinsic inefficiency is due to the dead space between counters. The
higher TSC inefficiency is due to the higher background
rate. The hit inefficiencies for MC and data agree to within 0.3\%.

\section{Result}
\label{sec:result}

\begin{figure}[th]
\centerline{
\epsfysize=3.5in
\epsfbox{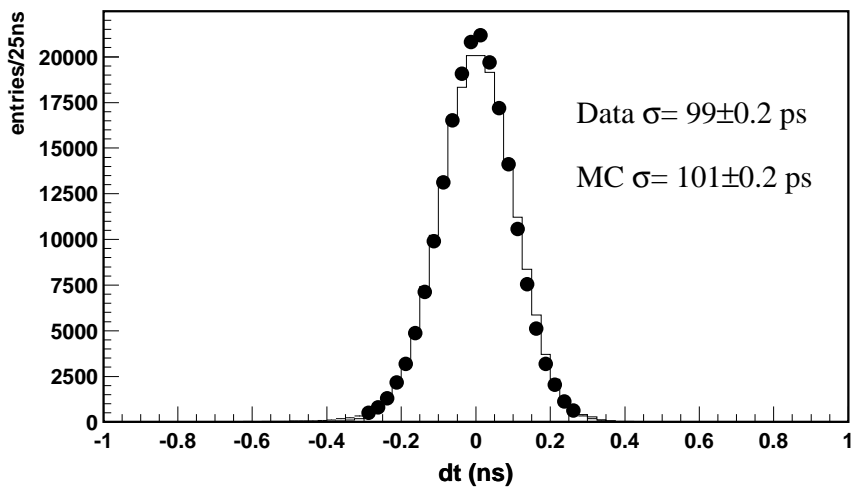}}
\caption{The $dt$ distribution. The circles are data; the histogram is MC. }
\label{fig:dt}
\end{figure}

\begin{figure}[ht]
\centerline{
\epsfysize=5.0in
\epsfbox{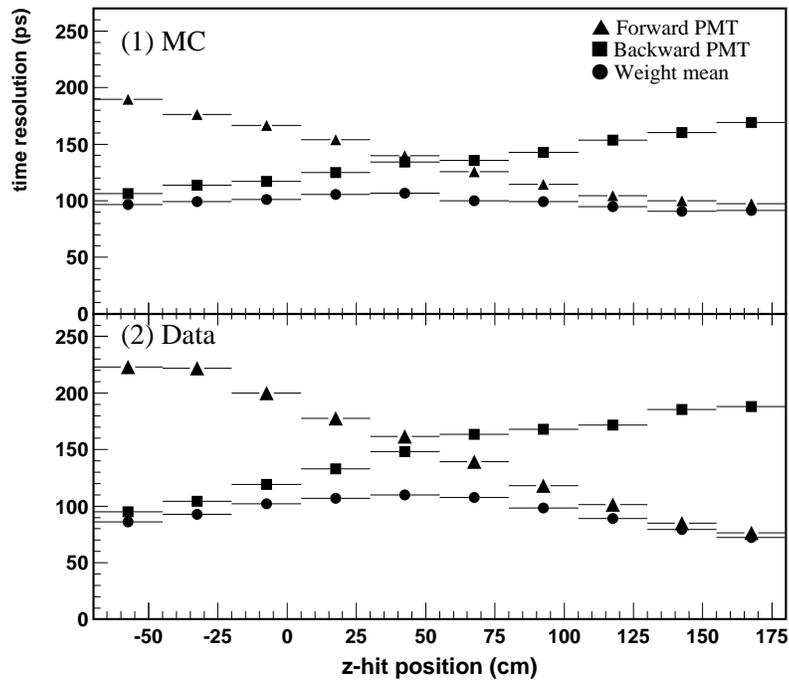}}
\caption{The TOF resolution as a function of $z$-hit. The top panel is
MC; the bottom is data.}
\label{fig:resolution}
\end{figure}

Figure~\ref{fig:dt} shows the $dt=T_{obs}^{correct}-T_{pred}$ distribution
for
all available TOF hits in a $\mu$-pair event sample. Here $T_{pred}$ is the time
of flight predicted using the track path length calculated from the fit
to hits in the CDC, and $T_{obs}^{correct}$ is the weighted
average of the observed times in the forward and backward PMTs after
correction for the $z$-hit position and time walk ~\cite{TOFNIM}. The distribution is fitted to a symmetric Gaussian to determine the time resolution. The
overall time resolution ($100~\rm{ps}$) is
well represented by the MC. Figure~\ref{fig:resolution} shows the time
resolutions for forward and backward PMTs and for the weighted average
time as a function of $z$-hit .

\begin{figure}[th]
\centerline{
\epsfysize=4.5in
\epsfbox{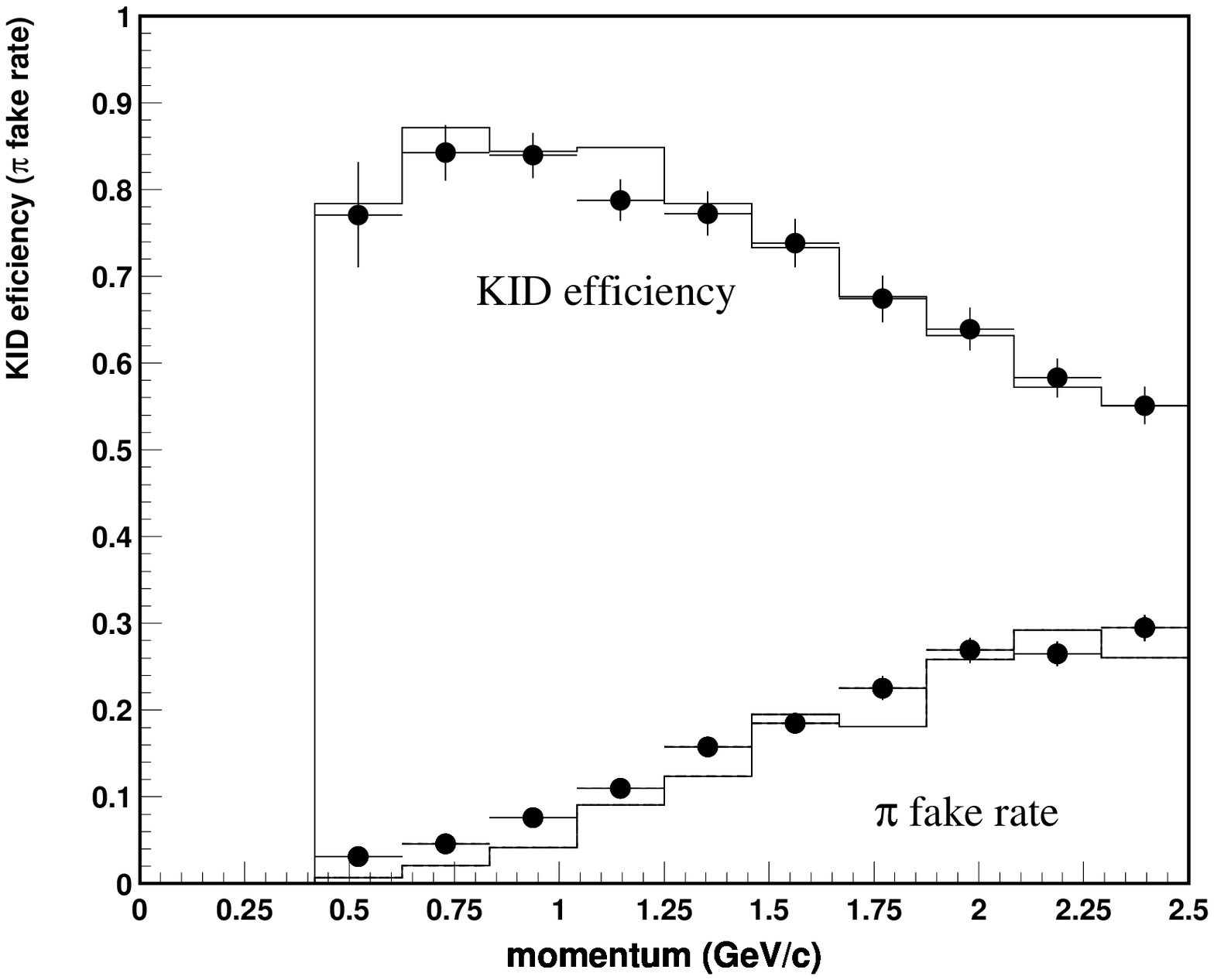}}
\caption{The $K$ ID efficiency and $\pi$ fake rate as a function of
momentum. The circles are data; the histogram is MC. }
\label{fig:kid}
\end{figure}

The PID from the Belle TOF system is based on a ratio of likelihoods for
two particle
hypotheses. For $K/\pi$ separation, a particle is identified as a kaon
using the likelihood ratio $PID(K) = {P_K}/(P_{\pi}+P_{K})$, where the
$P_{K,\pi}$ are the PID likelihoods for the particle hypotheses in the
subscripts. The PID performance is evaluated using the charm decay
$D^{*+} \rightarrow D^0\pi^+$, followed by  $D^0 \rightarrow K^- \pi^+$.
Figure~\ref{fig:kid}
shows the measured $K$ ID efficiency and $\pi$ fake rate as a function of
momentum, where $PID(K)> 0.6$ is required to identify a kaon. The MC
reproduces the $K$ ID efficiency and fake rate as functions of
momentum with maximum deviations of 0.7\% and 1.8\% , respectively.

\section{Conclusion}
\label{sec:conclusion}

In this report we describe the full detector simulation for the
Belle TOF system. We developed a GEANT-based simulator for the
processes in the scintillator and the PMT, the readout electronics of the
discriminator and the QTC, and the effect of beam background.

Scintillation light is produced using the hit information from
GEANT. Using simple models for light propagation in the TOF scintillator
and for the PMT response, we accurately reproduce the real PMT
signals. The simulation
represents the detector characteristics of the TOF system such as light
attenuation, the ADC distribution, the effective light velocity, the TDC
distributions and the time resolution. Through a beam background merging
process, the QTC dead time and the limit on the number of records in the
TDC1877
system is simulated, so that the TOF hit inefficiency dependence on the
background rate is accurately modeled.
Since the dependence on pulse height and on $z$-hit is well reproduced,
the simulation is a good tool for the investigation and development of
calibration and reconstruction methods.
The $K$ ID efficiency and its momentum dependence are in good
agreement with the data. The TOF simulation 
provides an important component of the Belle MC used for physics
analysis.

\section{Acknowledgements}

We wish to thank the Belle collaboration and KEKB accelerator group.
We acknowledge support from the
Ministry of Education, Culture, Sports, Science, and Technology of Japan
and the Japan Society for the Promotion of Science; the BK21 program of
the Ministry of Education of Korea and the Center for High Energy Physics
sponsored by the KOSEF; and the U.S. Department of Energy.

\clearpage

\begin{table}[ht]
\caption{Characteristic parameters of the TOF scintillator and the FM-PMT.}
\label{tb:TOFPMT}
\begin{center}
\begin{tabular}{c c}
\hline
\multicolumn{2} {c} {TOF scintillator (BC408)}  \\\hline
	     Base & Polyvinyltoluene\\
	     Density & 1.032\\
	     Refractive index & 1.58\\
	     Rise Time & 0.9 ns \\
	     Decay Time & 2.1 ns \\
 	     Pulse Width & $\sim$ 2.5 ns \\
             Light Attenuation Length &$\sim$ 300 cm \\
	     Wavelength of Maximum Emission & 425 nm \\\hline\hline

\multicolumn{2} {c} {FM PMT (R6680)}  \\\hline
	Effective photocathode diameter & 39 mm \\
	Transit Time Spread &  320 ps (r.m.s) \\
	Quantum Efficiency & $\sim$ 0.21\\
	Electron Collection Factor & 0.6\\
	Rise Time &  3.5 ns \\
	Fall Time & 4.5 ns \\
	Pulse width  & 6 ns (FWHM) \\
\hline
\end{tabular}
\end{center}
\end{table}



\begin{thebibliography}{99}
\bibitem{aging1} {After 30 months of data taking, we observed an
attenuation length 13\% smaller
than the initial value due to aging of the
scintillator. The
data shown in this figure are from the initial stages of the experiment.}
\end{thebibliography}
\end{document}